# A definition of the running coupling constant in a twisted SU(2) lattice gauge theory.

G.M. de Divitiis, R. Frezzotti, M. Guagnelli and R. Petronzio

Dipartimento di Fisica, Università di Roma *Tor Vergata*

and

INFN, Sezione di Roma II

Viale della Ricerca Scientifica, 00173 Roma, Italy

### Abstract

We propose a definition of the running coupling constant in a SU(2) lattice gauge theory with twisted boundary conditions. It is based on the correlation of Polyakov loops extended in a twisted direction at a distance which is a fixed fraction of the totale lattice size. We make the perturbative calculation which connects this definition to standard regularization schemes. We find $\Lambda_{\mathrm{Twisted-Polyakov}}/\Lambda_{\overline{\mathrm{MS}}} = 1.6136(2)$.

# 1 Introduction

The precise determination of the strong coupling constant $\alpha_s$ from experiments at intermediate energies suffers from systematic errors due to our ignorance of the next to leading terms of the perturbative expansion. At high energies, the perturbative estimates are more reliable, but even a relatively precise value obtained for $\alpha_s$ when translated into a value for $\Lambda$ leads to a rather approximate result. The reason is the asymptotically free character of the theory which amplifies the relative error on $\Lambda$ with respect to the one on $\alpha_s$ by a factor proportional to its inverse. Lattice calculations can in principle relate the determination of $\alpha_s$ to non perturbative quantities [1] with systematic errors which can be kept under control. The ideal calculation would be the reconstruction of the running coupling constant from a low energy scale normalized by a non perturbative quantity – the proton mass, for example – to a high energy scale where the connection to a standard regularization scheme like $\overline{\mathrm{MS}}$ can be safely performed within perturbation theory. The highest energy scale available on the lattice is provided by the momentum cutoff $\pi/a$ where $a$ is the lattice spacing.

However, if one wants to reconcile a sizeable total volume with a perturbative scale at distances of the order of the lattice spacing he needs a number of lattice points exceeding the computing power available in a near future. A finite size renormalization group technique can solve this problem: this method has been widely used in the study of critical points of statistical systems [2] and in the context of the determination of $\alpha_s$ by the authors of references [3, 4, 5]. The coupling constant running from high to low energies is constructed through the following steps:

- one defines a finite volume coupling constant renormalized at a scale depending upon the total volume. Its relation to the lattice bare coupling constant should be well defined and calculable in the continuum limit.

- one calculates the variation of the value of the coupling between a volume with $N_1$ points per side (the "small" volume) and one with $N_2 = 2N_1$ points (the "large" volume). This calculation is repeated for several lattice spacing at fixed physical volumes and extrapolated to the zero lattice spacing limit.

- in order to proceed to larger physical volumes, one needs a readjustment of the lattice spacing. This is done by matching the value of the renormalized coupling on a "large" volume with the one calculated on a "small" volume at a different value of the bare lattice coupling.

- the last two steps are iterated until the total physical volume is large enough to guarantee the absence of finite size effects in the non perturbative calculation which will normalize the volume in physical units.

In this paper we present in the case of SU(2) with twisted boundary conditions a perturbative calculation which connects a definition of $\alpha_s$ based on the correlation of Polyakov loops with the $\overline{\mathrm{MS}}$ scheme. Section 2 is devoted to a short discussion of twisted boundary conditions and of the proper definition of a twisted Polyakov loop correlation. Section 3 presents the details of the calculation and a comment on the absence of linear divergences in the observable used for defining $\alpha_s$. The last section contains the final result for the ratio $\Lambda_{\overline{\mathrm{MS}}}/\Lambda_{\mathrm{Twisted-Polyakov}}$ where $\Lambda_{\mathrm{Twisted-Polyakov}}$ ($\equiv \Lambda_{\mathrm{TP}}$ in the following) is the $\Lambda$ parameter of our scheme.

## 2 Twisted boundary conditions

The method proposed relies on a definition of $\alpha_s$ at a scale fixed by the total finite volume. The perturbative calculation which connects this definition to a standard one should then be performed on a finite volume. This computation cannot be done in a straightforward way because of the existence of gauge inequivalent field configurations of degenerate minimal energy: the torons [9]. They are parametrized by constant and diagonal link matrices which give the same contribution to the lattice action as the unity matrices which parametrize the standard perturbative vacuum. A solution to this problem is to adopt twisted boundary conditions at least in two directions [7, 8] which make the toron contribution to the energy higher than the vacuum one.

Normal periodic boundary conditions can be seen as the surrounding of the original lattice with replicas in all directions. Twisted boundary conditions define the configuration in a replica which lies in a twisted direction as the original one after a particular and in our case global gauge transformation.

By calling $\Omega_\nu$ the SU($N$) matrix corresponding to the twist in the $\nu$ direction and $U_\mu(r)$ the link variable connecting the site $r$ to the site $r+\hat{\mu}$ (where $r=(x,y,z,t)$), a minimal twisting condition reads:

$$
\begin{aligned}
U_\mu(r+\hat{\nu}L) &= \Omega_\nu U_\mu(r)\Omega_\nu^\dagger \qquad && \hat{\nu}=\hat{x},\hat{y} \\
U_\mu(r+\hat{\nu}L) &= U_\mu(r) \qquad && \hat{\nu}=\hat{z},\hat{t}
\end{aligned}
\tag{1}
$$

with $U_\mu(r+\hat{\nu}L)$ a "replica–link", *i.e.* a link in a site displaced by the lattice size in the $\nu$ direction and $\Omega_x$, $\Omega_y$ two non commuting unitary matrices. Throughout the paper we will adopt lattice spacing units ($a=1$). If the link variables $U_\mu(r) \in$ SU($N$) are parametrized by:

$$
\begin{aligned}
U_\mu(r) &= \exp\left(i g_0 A_\mu(r)\right) \qquad && \text{with} \\
A_\mu^\dagger(r) &= A_\mu(r) \qquad && \mathrm{Tr}(A_\mu(r)) = 0
\end{aligned}
\tag{2}
$$

the boundary condition (1) implies for gauge fields $A_\mu(r)$:

$$
\begin{aligned}
A_\mu(r+\hat{\nu}L) &= \Omega_\nu A_\mu(r)\Omega_\nu^\dagger \qquad && \hat{\nu}=\hat{x},\hat{y} \\
A_\mu(r+\hat{\nu}L) &= A_\mu(r) \qquad && \hat{\nu}=\hat{z},\hat{t}
\end{aligned}
\tag{3}
$$

The value of the link variable in a position which is reached after two independent twisted translations cannot depend upon the order in which they are performed. This is realized if the $\Omega_\mu$ satisfy:

$$
\Omega_x\Omega_y = \mathbf{z}\Omega_y\Omega_x \tag{4}
$$

where $\mathbf{z}=\exp(2\pi ik/N)$, $k=1,\ldots,N-1$; in our SU(2) case $k=1$.

After a combined translation the transformed link variable is:

$$
U_\mu(r+\hat{x}L+\hat{y}L) = \Omega_x\Omega_y U_\mu(r)\Omega_y^\dagger\Omega_x^\dagger \tag{5}
$$

If the condition (4) is satisfied, exchanging the order of the translations corresponds to interchanging $\Omega_x$ and $\Omega_y$ in expression (5) and leads to the same result. The gauge transformation for the original links

$$
U_\mu(r) \quad \rightarrow \quad \Lambda(r)U_\mu(r)\Lambda^\dagger(r+\hat{\mu}) \tag{6}
$$

implies that the replica–links transform with a twisted gauge matrix:

$$
\Lambda(r+\hat{\nu}L) = \Omega_\nu\Lambda(r)\Omega_\nu^\dagger \tag{7}
$$

In the standard Wilson action with twisted boundary conditions the plaquettes at the boundary of the twisted directions contain a replica–link. For example, at the boundary $y = L-1$, the expression of a plaquette in the $x$–$y$ plane in terms of the original link variables is given by:

$$U_P^{xy} = U_x(x, L-1, z, t) U_y(x+1, L-1, z, t) \times \\ \Omega_y U_x^\dagger(x+1, 0, z, t) \Omega_y^\dagger U_y^\dagger(x, L-1, z, t) \,. \tag{8}$$

Toron configurations are no longer degenerate in energy with the usual vacuum because they are not transformed into themselves by the twisting conditions and give a contribution to the boundary plaquettes which is different from the one of the standard vacuum.

Our definition of $\alpha_s$ is in terms of the correlation of Polyakov loops. These are defined as the trace of the product of links in a fixed direction and with the same position in the hyperplane orthogonal to that direction. In order to obtain a gauge invariant operator, if the direction of the Polyakov loop is a twisted one, one has to trace with the corresponding twist matrix:

$$P_x(y, z, t) = \mathrm{Tr}([\prod_{j=0}^{L-1} U_x(x = j, y, z, t)]\Omega_x) \tag{9}$$

The correlation of two twisted Polyakov loops starts at order $g_0^2$ and is a natural candidate for defining $\alpha_s$. We take the following quantity:

$$\mathcal{O} = \frac{\langle \sum_{y,z} P_x(0,0,0) P_x(y, z, t = L/2) e^{i\frac{\pi}{L}y} \rangle}{\langle \sum_{x,y} P_z(0,0,0) P_z(x, y, t = L/2) \rangle} \tag{10}$$

where the numerator and the denominator are the ground state expectation values of the correlations of twisted and normal Polyakov loops respectively. The phase multiplying the correlation ensures the invariance under translations in all twisted directions.

The numerator and the denominator of equation (10) are affected by linear divergences which spoil their continuum limit. Various authors [10] have argued that these divergences do exponentiate and factorize; therefore the ratio of eq. (10) is a well defined quantity affected only by logarithmic divergences which can be reabsorbed in the renormalized coupling constant. In the next section we explicitely verify this statement up to order $g_0^4$.

We define the renormalized running coupling at the scale $L$, $\overline{g}^2(L)$, through:

$$\mathcal{O}(L) \equiv T(L)\overline{g}^2(L) \tag{11}$$

where $T(L)$ is the lowest tree–level contribution to $\mathcal{O}$.

# 3 The SU(2) perturbative calculation

We start from the SU(2) pure gauge action:

$$S = \frac{2}{g_0^2} \sum_p \mathrm{Tr}(1 - U_p) \tag{12}$$

where we remind that when the links are in the replica they are replica–links satisfying the twisted boundary conditions (1).

The connection of the definition in eq. (11) to other regularization schemes requires the calculation of the one–loop correction to the observable $\mathcal{O}$, *i.e.* of terms up to order $g_0^4$. These come from the expansion of the action up to order $g_0^2$ and of the operator up to order $g_0^4$. We have performed our calculation in configuration space to which refer our Feynman rules. The explicit form of the various terms defined by:

$$S = S_0 + g_0^2 S_{\mathrm{meas}} + g_0 S_{\mathrm{W}}^{(1)} + g_0^2 S_{\mathrm{W}}^{(2)} + g_0 S_{\mathrm{ghost}}^{(1)} + g_0^2 S_{\mathrm{ghost}}^{(2)} + \mathrm{O}(g_0^3) \tag{13}$$

is given in Appendix A. The Feynman diagrams are represented in figure 1 and 2.

The basic building block is the gluon propagator which, due to the twisted boundary conditions, has peculiar properties which depend upon the explicit form chosen for the $\Omega$ matrices. For our SU(2) case, the non trivial $\Omega$'s can be identified with the Pauli matrices:

$$\Omega_x = i\sigma_1 \qquad\qquad \Omega_y = i\sigma_2 \tag{14}$$

If the gauge fields of the SU(2) algebra are expressed in terms of the Pauli matrices $\sigma_a$ (a=1,2,3):

$$A_\mu(r) = \sum_{a=1}^{3} A_\mu^a(r)\, \sigma_a \tag{15}$$

equation (3) reads:

$$A^a_\mu(r+\hat{\nu}L) = \frac{1}{2}\mathrm{Tr}(\sigma_a\Omega_\nu\sigma_b\Omega^\dagger_\nu)A^b_\mu(r) \qquad \hat{\nu}=\hat{x},\hat{y}$$
$$A^a_\mu(r+\hat{\nu}L) = A^a_\mu(r) \qquad \hat{\nu}=\hat{z},\hat{t} \tag{16}$$

which, together with eq. (14), implies that the gauge fields of colour 1 are periodic in the direction $y$ over an anomalously enlarged space of size $2\times L$. The same happens for those of colour 2 in the direction $x$ and those of colour 3 in both $x$ and $y$ directions. The propagator is diagonal in colour and shares the periodicity properties discussed above as a function of the distance between two points.

Equation (14) associates to each twist matrix a corresponding colour: in our basis the one corresponding to translations in direction $x$ ($y$) is associated to colour 1(2). One can define new vector potentials which have a standard periodicity in all directions. They are related to the old ones by a local colour–depending phase, according to:

$$\tilde{A}^a_\mu(r) \;=\; A^a_\mu(r)\;\exp(-i\frac{\pi}{L}(n_a\cdot r)) \qquad \text{with}$$
$$n_1=(0,1,0,0) \qquad n_2=(1,0,0,0) \qquad n_3=(1,1,0,0) \tag{17}$$

Operators built from links expressed in terms of the new vector potentials have normal translation invariance properties. The phase appearing in the definition of the correlation between twisted Polyakov loops is just the net result of the transformation from a loop expressed in terms of the new links to one in terms of the old. It can also be verified explicitely that tracing a single twisted loop with the appropriate twist matrix times a phase corresponding to the associated colour produces an ordinary translation invariant operator.

The calculation in configuration space was done using FORM [12], an algebraic manipulator, in order to deal with the plethora of Wick contractions. The sums over the positions were done with a numerical FORTRAN program. Their number in the most time consuming diagram was reduced from twelve to five. The gluon and ghost propagators in configuration space were obtained by doing a fast Fourier transform of their analytic expression in momentum space (see Appendix B for details on the momentum space formulation).

# 4 The results

Our calculation passed successfully some crucial tests.

(i) The diagrams contributing to the gluon self energy exhibit ultraviolet quadratic divergences which appear as contributions growing quadratically with the number of lattice points per side. These divergences cancel.

(ii) The sum of diagrams of the twisted Polyakov loops correlation has genuine linear divergences which cancel against the product of the denominator at order $g_0^2$ times the tree level numerator.

(iii) After the above cancellations one is left with the logarithmic divergences which make the renormalized $\alpha_s$ running with the lattice size. Their coefficient $s_0$ is known from continuum calculations and is given by:

$$\frac{11}{12\pi^2} \equiv 2\beta_0 \equiv s_0^{\mathrm{th}} \simeq 0.092878 \tag{18}$$

to be compared with our result:

$$s_0 = 0.092880 \pm 0.000005 \tag{19}$$

In the following we shall adopt this notation:

$$\mathcal{O}(L) = T(L)[g_0^2 + g_0^4 D(L)] \tag{20}$$

where $D(L)$ is the (tree–level normalized) one–loop contribution.

The $L$ dependence of the one–loop contribution can be parametrized as[1]:

$$D(L) = r_0 + s_0 \log(L) + \frac{r_1}{L} + \frac{r_2}{L^2} + \frac{s_2}{L^2}\log(L) + \ldots \tag{21}$$

[1]A term of the form $\frac{s_1}{L}\log(L)$ is ruled out: the lattice pure gauge action differs from the continuum one only by terms of order $a^2$, *i.e.* in our units of order $1/L^2$, which implies that terms of the above form cannot be generated at one loop level. The $r_1/L$ term could be produced by combining linear divergences with $1/L^2$ corrections. However, it turns out to be absent because the linear divergences appearing in the numerator and denominator of eq. (10), which defines $\mathcal{O}$, fully factorize and exactly cancel in the ratio.

| $L$–range | $r_0$ | $s_0$ | $r_1$ |
|---|---|---|---|
| 4–32 | 0.2897 | 0.1012 | 0.1349 |
| 8–32 | 0.3160 | $9.430 \cdot 10^{-2}$ | $3.061 \cdot 10^{-2}$ |
| 12–32 | 0.3194 | $9.345 \cdot 10^{-2}$ | $1.388 \cdot 10^{-2}$ |
| 16–32 | 0.3206 | $9.316 \cdot 10^{-2}$ | $7.325 \cdot 10^{-3}$ |
| 18–32 | 0.3211 | $9.304 \cdot 10^{-2}$ | $4.381 \cdot 10^{-3}$ |
| 20–32 | 0.3213 | $9.300 \cdot 10^{-2}$ | $3.476 \cdot 10^{-3}$ |
| 22–32 | 0.3219 | $9.286 \cdot 10^{-2}$ | $-5.167 \cdot 10^{-4}$ |

Table 1: Results of the first fit showing numerical evidence for $r_1 = 0$

Various fits of $D(L)$ were made, for a series of lattices ranging from $L = 4$ to $L = 32$.

The first was addressed to obtain a numerical evidence for the absence of the $1/L$ terms; in table 1 we show the results of the fit:

$$D(L) = r_0 + s_0 \log(L) + \frac{r_1}{L} \tag{22}$$

for different ranges of lattice sizes. While the fitted values of the coefficients of the logarithm $s_0$ and of the constant $r_0$ show no appreciable dependence upon the smallest lattice size $L$ used in the fit, the value of the linear term $r_1$ decreases systematically, providing evidence for the absence of such a contribution.

Assuming that $r_1 = 0$, the second set of fits (see table 2) was obtained by fitting $D(L)$ with the form $r_0 + s_0 \log(L)$ and led to a precise determination of the coefficient of the logarithm with the result in eq. (19).

The precision ($\simeq 5 \times 10^{-6}$) of this determination of $r_0$ is better than what we expected: neglecting in the expression for $D(L)$ terms of the form $L^{-2}$ and $L^{-2} \log(L)$ induces an error, which, in principle, should be of order $10^{-3}$, for the range of lattice sizes of interest. In order to clarify this point, we made (for various ranges of lattice sizes) various fits of the quantity $D(L) - s_0^{\mathrm{th}} \log(L)$ with the expression:

$$r_0 + \frac{r_2}{L^2} + \frac{s_2}{L^2} \log(L) \; . \tag{23}$$

In the range of lattice sizes $22 \leq L \leq 32$, which leads to a stable determination of $r_0$ and $s_0$ in the previous set of fits (see table 2), we found for $r_2$ and $s_2$ values that make the combination $(r_2 + s_2 \log(L))/L^2$ always smaller than $10^{-5}$, with the value for $r_0$ stable within a precision of $10^{-5}$. In this range of lattice sizes we are allowed to neglect in our fits the terms of order $L^{-2}$ and $L^{-2}\log(L)$ and we obtain results with a precision of some units in $10^{-5}$.

| $L$–range | $r_0$ | $s_0$ | $\delta_{\rm max}$ |
|---|---|---|---|
| 16–32 | 0.32199 | $9.2833 \cdot 10^{-2}$ | $2.8 \cdot 10^{-5}$ |
| 18–32 | 0.32191 | $9.2857 \cdot 10^{-2}$ | $1.2 \cdot 10^{-5}$ |
| 20–32 | 0.32188 | $9.2868 \cdot 10^{-2}$ | $1.0 \cdot 10^{-5}$ |
| 22–32 | 0.32184 | $9.2878 \cdot 10^{-2}$ | $9.2 \cdot 10^{-6}$ |
| 24–32 | 0.32183 | $9.2883 \cdot 10^{-2}$ | $9.0 \cdot 10^{-6}$ |
| 26–32 | 0.32184 | $9.2878 \cdot 10^{-2}$ | $9.0 \cdot 10^{-6}$ |

Table 2: Results of the second fit determining the value of $s_0$, where $\delta_{\rm max} = {\rm Max}_{\{L\}} \left( |D(L) - s_0 \log(L) - r_0| / |D(L)| \right)$

The last set of fits was made with the coefficient of the purely logarithmic term fixed to its theoretical value, $s_0^{\rm th}$, to get the best determination of the constant term which fixes the renormalization scheme (see table 3 and figure 3). For various ranges of lattice sizes, we made a fit of the quantity

$$D(L) - s_0^{\rm th} \log(L) \tag{24}$$

with a single constant parameter ($r_0$).
Our final result is:

$$r_0 = 0.32185(1) \tag{25}$$

where the error was estimated by $\epsilon_{\rm max} r_0 \simeq 10^{-5}$ ($\epsilon_{\rm max}$ is defined in table 3).

This constant settles the relation to any other regularization scheme and allows a comparison of the numerical result for the running coupling constant with other definitions. The ratio $\Lambda_{\rm TP}/\Lambda_{\overline{\rm MS}}$ gives the relationship between our definition of $\alpha_s$ and the one of the $\overline{\rm MS}$ scheme.

We make use of the relation [11] between the coupling constant in the $\overline{\rm MS}$ regularization scheme, $g^2_{\overline{\rm MS}}$, and the lattice renormalized coupling, $g^2_{\rm latt}$, defined by

$$g^2_{\rm latt}(\mu^{-1}) = \frac{1}{2\beta_0 \log(\mu/\Lambda_{\rm latt})} \quad , \tag{26}$$

| $L$–range | $r_0$ | $\epsilon_{\max}$ |
|---|---|---|
| 12–32 | 0.32188 | $4.8 \cdot 10^{-4}$ |
| 16–32 | 0.32185 | $1.0 \cdot 10^{-4}$ |
| 18–32 | 0.32185 | $3.6 \cdot 10^{-5}$ |
| 20–32 | 0.32185 | $1.9 \cdot 10^{-5}$ |
| 22–32 | 0.32185 | $1.9 \cdot 10^{-5}$ |
| 24–32 | 0.32185 | $1.9 \cdot 10^{-5}$ |
| 26–32 | 0.32185 | $1.8 \cdot 10^{-5}$ |
| 28–32 | 0.32185 | $1.6 \cdot 10^{-5}$ |

Table 3: Final fit for $r_0$, where $\epsilon_{\max} = \mathrm{Max}_{\{L\}}\left(|D(L) - s_0^{\mathrm{th}}\log(L) - r_0|/|r_0|\right)$

where $g^2_{\mathrm{latt}}(1) \equiv g_0^2$. The knowledge of $r_0$ fixes the relation between $g^2_{\mathrm{TP}}$ and $g^2_{\mathrm{latt}}$, which reads, to one–loop order:

$$g^2_{\mathrm{TP}}(\mu^{-1}) \equiv \frac{1}{2\beta_0 \log(\mu/\Lambda_{\mathrm{TP}})} = g^2_{\mathrm{latt}}(\mu^{-1}) + r_0 g^4_{\mathrm{latt}}(\mu^{-1}) \tag{27}$$

where $\mu^{-1}$ is a physical length. The desired relationship is given by:

$$\frac{\Lambda_{\mathrm{TP}}}{\Lambda_{\overline{\mathrm{MS}}}} = \frac{\Lambda_{\mathrm{latt}}}{\Lambda_{\overline{\mathrm{MS}}}} \exp(\frac{r_0}{2\beta_0}) = 1.6136(2) \quad . \tag{28}$$

# Appendix A

We have made the perturbative calculation of the observable $\mathcal{O}$ in eq.(10) up to the order $g_0^4$ for the expectation value in the numerator and $g_0^2$ for the one in the denominator. The expectation values have the standard definition in terms of ratios of functional integrals:

$$\langle \mathcal{C} \rangle = \frac{\int DU e^{-S_W[U]} \mathcal{C}[U]}{\int DU e^{-S_W[U]}} \tag{29}$$

After the Feynman gauge fixing, the functional integral becomes:

$$\langle \mathcal{C} \rangle = \frac{\int D\bar{c} Dc DA e^{-S[A,\bar{c},c]} \mathcal{C}[A]}{\int D\bar{c} Dc DA e^{-S[A,\bar{c},c]}} \tag{30}$$

where $c, \bar{c}$ are the ghost fields and $S$ is given by

$$S = S_0 + g_0^2 S_{meas} + g_0 S_W^{(1)} + g_0^2 S_W^{(2)} + + g_0 S_{ghost}^{(1)} + g_0^2 S_{ghost}^{(2)} + O(g_0^3) \quad (31)$$

The Feynman diagrams needed to perform the calculation are given for the numerator of eq.(10) in figure 1 and for denominator in figure 2.
Diagrams A, B and C need the expansion of the twisted Polyakov loop up to order $g_0^3$. With the exception of diagram I which is zero, the remaining ones all need the twisted Polyakov loop only to order $g_0$. Moreover, diagrams D and E need the gauge field action up to order $g_0^2$, diagrams F contains the measure at order $g_0^2$, and diagrams G and H involve the ghost action up to order $g_0^2$. Diagrams in figure 2 involve only the Polyakov loop at order $g_0^2$.

The expansion of the operator is straightforward. The one of the gauge action is a little bit more involved and we report in the following the explicit expressions for the quantities of eq.(31).

$$S_W^{(1)} = -4 \sum_r \sum_{\mu\neq\nu} \{ \vec{A}_\mu(r) \times \vec{A}_\mu(r+\hat{\nu}) \cdot [\vec{A}_\nu(r+\hat{\mu}) + \vec{A}_\nu(r)] \} \quad (32)$$

$$\begin{aligned}
S_W^{(2)} = -\tfrac{1}{6} \sum_r \sum_{\mu\neq\nu} \{ & +4\, [\vec{A}_\nu(r) \cdot \vec{A}_\mu(r+\hat{\nu})]\ [\vec{A}_\nu(r) \cdot \vec{A}_\nu(r)] + \\
& +4\, [\vec{A}_\nu(r) \cdot \vec{A}_\mu(r+\hat{\nu})]\ [\vec{A}_\mu(r+\hat{\nu}) \cdot \vec{A}_\mu(r+\hat{\nu})] + \\
& -4\, [\vec{A}_\nu(r) \cdot \vec{A}_\nu(r+\hat{\mu})]\ [\vec{A}_\nu(r) \cdot \vec{A}_\nu(r)] + \\
& -4\, [\vec{A}_\nu(r) \cdot \vec{A}_\nu(r+\hat{\mu})]\ [\vec{A}_\nu(r+\hat{\mu}) \cdot \vec{A}_\nu(r+\hat{\mu})] + \\
& -2\, [\vec{A}_\nu(r) \cdot \vec{A}_\mu(r)]\ [\vec{A}_\nu(r) \cdot \vec{A}_\nu(r)] + \\
& -2\, [\vec{A}_\nu(r) \cdot \vec{A}_\mu(r)]\ [\vec{A}_\mu(r) \cdot \vec{A}_\mu(r)] + \\
& -2\, [\vec{A}_\nu(r+\hat{\mu}) \cdot \vec{A}_\mu(r+\hat{\nu})]\ [\vec{A}_\mu(r+\hat{\nu}) \cdot \vec{A}_\mu(r+\hat{\nu})] + \\
& -2\, [\vec{A}_\nu(r+\hat{\mu}) \cdot \vec{A}_\mu(r+\hat{\nu})]\ [\vec{A}_\nu(r+\hat{\mu}) \cdot \vec{A}_\nu(r+\hat{\mu})] + \\
& +6\, [\vec{A}_\nu(r) \cdot \vec{A}_\nu(r)]\ [\vec{A}_\mu(r+\hat{\nu}) \cdot \vec{A}_\mu(r+\hat{\nu})] + \\
& +6\, [\vec{A}_\nu(r) \cdot \vec{A}_\nu(r)]\ [\vec{A}_\nu(r+\hat{\mu}) \cdot \vec{A}_\nu(r+\hat{\mu})] + \\
& +3\, [\vec{A}_\nu(r) \cdot \vec{A}_\nu(r)]\ [\vec{A}_\mu(r) \cdot \vec{A}_\mu(r)] + \\
& +3\, [\vec{A}_\mu(r+\hat{\nu}) \cdot \vec{A}_\mu(r+\hat{\nu})]\ [\vec{A}_\nu(r+\hat{\mu}) \cdot \vec{A}_\nu(r+\hat{\mu})] + \\
& +\ [\vec{A}_\nu(r) \cdot \vec{A}_\nu(r)]\ [\vec{A}_\nu(r) \cdot \vec{A}_\nu(r)] + \\
& +\ [\vec{A}_\nu(r+\hat{\mu}) \cdot \vec{A}_\nu(r+\hat{\mu})]\ [\vec{A}_\nu(r+\hat{\mu}) \cdot \vec{A}_\nu(r+\hat{\mu})] +
\end{aligned}$$

$$
\begin{aligned}
&+12[\vec{A}_\nu(r)\cdot\vec{A}_\mu(r+\hat{\nu})]\ \ [\vec{A}_\nu(r+\hat{\mu})\cdot\vec{A}_\mu(r)]+\\
&+12[\vec{A}_\nu(r)\cdot\vec{A}_\mu(r)]\ \ [\vec{A}_\mu(r+\hat{\nu})\cdot\vec{A}_\nu(r+\hat{\mu})]+\\
&-12[\vec{A}_\nu(r)\cdot\vec{A}_\nu(r+\hat{\mu})]\ \ [\vec{A}_\mu(r+\hat{\nu})\cdot\vec{A}_\mu(r)]+\\
&+12[\vec{A}_\nu(r+\hat{\mu})\cdot\vec{A}_\mu(r)]\ \ [\vec{A}_\nu(r)\cdot\vec{A}_\nu(r)]+\\
&+12[\vec{A}_\nu(r+\hat{\mu})\cdot\vec{A}_\mu(r)]\ \ [\vec{A}_\mu(r+\hat{\nu})\cdot\vec{A}_\mu(r+\hat{\nu})]+\\
&-12[\vec{A}_\mu(r+\hat{\nu})\cdot\vec{A}_\nu(r+\hat{\mu})]\ \ [\vec{A}_\nu(r)\cdot\vec{A}_\nu(r)]+\\
&-12[\vec{A}_\mu(r+\hat{\nu})\cdot\vec{A}_\mu(r)]\ \ [\vec{A}_\nu(r)\cdot\vec{A}_\nu(r)]+\\
&-12[\vec{A}_\nu(r)\cdot\vec{A}_\nu(r+\hat{\mu})]\ \ [\vec{A}_\mu(r+\hat{\nu})\cdot\vec{A}_\mu(r+\hat{\nu})]+\\
&-12[\vec{A}_\nu(r)\cdot\vec{A}_\mu(r)]\ \ [\vec{A}_\mu(r+\hat{\nu})\cdot\vec{A}_\mu(r+\hat{\nu})]\}
\end{aligned}
\tag{33}
$$

$$
S_{meas} = \tfrac{1}{3}\sum_r\sum_\mu \quad \vec{A}_\mu(r)\cdot\vec{A}_\mu(r) \tag{34}
$$

$$
\begin{aligned}
S^{(1)}_{ghost} = 4\sum_r\sum_\mu\{ \quad &+\vec{\bar{c}}(r)\cdot[\vec{A}_\mu(r)\times(\vec{c}(r)+\vec{c}(r+\hat{\mu}))]\\
&-\vec{\bar{c}}(r)\cdot[\vec{A}_\mu(r-\hat{\mu})\times(\vec{c}(r-\hat{\mu})+\vec{c}(r))]\}
\end{aligned}
\tag{35}
$$

$$
\begin{aligned}
S^{(2)}_{ghost} = -\tfrac{4}{3}\sum_r\sum_\mu\{ \quad &+[\vec{\bar{c}}(r)\cdot(\vec{c}(r)-\vec{c}(r+\hat{\mu}))][\vec{A}_\mu(r)\cdot\vec{A}_\mu(r)]\\
&-[\vec{\bar{c}}(r)\cdot\vec{A}_\mu(r)][\vec{A}_\mu(r)\cdot(\vec{c}(r)-\vec{c}(r+\hat{\mu}))]\\
&-[\vec{\bar{c}}(r)\cdot(\vec{c}(r-\hat{\mu})-\vec{c}(r))][\vec{A}_\mu(r-\hat{\mu})\cdot\vec{A}_\mu(r-\hat{\mu})]\\
&+[\vec{\bar{c}}(r)\cdot\vec{A}_\mu(r-\hat{\mu})][\vec{A}_\mu(r-\hat{\mu})\cdot(\vec{c}(r-\hat{\mu})-\vec{c}(r))]\}
\end{aligned}
\tag{36}
$$

Vectors are defined in the colour space and scalar and vector products accordingly:

$$
\begin{aligned}
&\vec{A} = (A^{(1)}, A^{(2)}, A^{(3)})\\
&\vec{c} = (c^{(1)}, c^{(2)}, c^{(3)})\\
&\vec{A}\cdot\vec{c} = A^{(1)}c^{(1)} + A^{(2)}c^{(2)} + A^{(3)}c^{(3)}\\
&(\vec{A}\times\vec{c})_i = \epsilon_{ijk}A^{(j)}c^{(k)}
\end{aligned}
$$

The sums $\sum_r$ range over all the points of the original lattice. This implies that some terms contain replica vector fields which are related to the original ones by eq.(16).

# Appendix B

In this appendix we want to discuss the relation between the colour degrees of freedom and the extra momentum degrees of freedom introduced on the lattice by the twisted boundary conditions (see equations (1) and (3)).

Expanding the SU($N$) gauge field $A_\mu(r)$ into plane waves

$$A_\mu(r) = \frac{1}{L^4}\sum_k \Gamma_k \tilde{A}_\mu(k) \mathrm{e}^{ikr + i\frac{k_\mu}{2}} \tag{37}$$

the twisting condition (3) implies:

$$\Omega_\nu \Gamma_k \Omega_\nu^\dagger = \mathrm{e}^{ik_\nu L}\Gamma_k \qquad \hat{\nu} = \hat{x}, \hat{y} \tag{38}$$

where $\Gamma_k$ is a complex $N \times N$ matrix. It can be shown (see ref. [6]) that a non–zero solution of equation (38) exists if and only if the momentum components in the twisted directions ($\hat{\nu} = \hat{x}, \hat{y}$) satisfy

$$\begin{aligned} k_\nu &= k_\nu^{\mathrm{ph}} + k_\nu^{\perp} && \text{with} \\ k_\nu^{\mathrm{ph}} &= \frac{2\pi}{L} n_\nu^{\mathrm{ph}} && -\frac{L}{2} \le n_\nu^{\mathrm{ph}} < \frac{L}{2} \\ k_\nu^{\perp} &= \frac{2\pi}{NL} n_\nu^{\perp} && n_\nu^{\perp} = 0, 1, \ldots, N-1 \end{aligned} \tag{39}$$

These momentum components, as expected, are quantized, but the $k_\perp$ term in the momentum adds to the ordinary momentum degrees of freedom those of a lattice $N$ times larger in each twisted direction. In non–twisted directions ($\hat{\nu} = \hat{z}, \hat{t}$) one can think of $k_\nu$ as a sum $k_\nu^{\mathrm{ph}} + k_\nu^{\perp}$, with $k_\nu^{\perp} = 0$, because in these directions the lattice of size $L$ exhibits the ordinary periodicity.

Up to an overall arbitrary phase, the solution of equation (38) is unique and reads

$$\Gamma_k = \Omega_x^{-n_y^\perp} \Omega_y^{n_x^\perp} \mathbf{z}^{\frac{1}{2}(n_x^\perp + n_y^\perp + 1)(n_x^\perp + n_y^\perp)} \tag{40}$$

where we used the $\Omega$ property

$$\Omega_\nu^N = (-1)^{N-1}\mathbf{1} \qquad \hat{\nu} = \hat{x}, \hat{y}. \tag{41}$$

From eq.(40) follows that the $\Gamma_k$'s are SU($N$) matrices, that they depend only upon $k_\perp$ and that [5] their normalization is:

$$\frac{1}{N}\mathrm{Tr}(\Gamma^\dagger_{k^\perp}\Gamma_{\tilde{k}^\perp}) = \delta^{(2)}_{k^\perp,\tilde{k}^\perp} \tag{42}$$

where $\delta^{(2)}_{k^\perp,\tilde{k}^\perp}$ is 1 if, for $\hat{\nu} = \hat{x}, \hat{y}$, $n^\perp_\nu = \tilde{n}^\perp_\nu (\mathrm{mod}\ N)$ and zero otherwise.

Equation (37) can be rewritten as

$$A_\mu(r) = \frac{1}{L^4}\sum_{k^{\mathrm{ph}},k^\perp} \Gamma_{k^\perp}\tilde{A}_\mu(k^{\mathrm{ph}},k^\perp)\mathrm{e}^{ikr+i\frac{k_\mu}{2}} \tag{43}$$

where the condition $\mathrm{Tr}(A_\mu(r)) = 0$ is implemented by requiring

$$\tilde{A}_\mu(k^{\mathrm{ph}},k^\perp) = 0 \qquad \text{if} \quad n^\perp_\nu = 0(\mathrm{mod}\ N) \quad . \tag{44}$$

The gluon propagator, in momentum space and in the Feynman gauge[2], is:

$$\langle \tilde{A}_\mu(q^{\mathrm{ph}},q^\perp)\tilde{A}_\nu(k^{\mathrm{ph}},k^\perp)\rangle = \frac{1}{2N}\delta^{(4)}_{(q+k)^{\mathrm{ph}},0}\delta^{(2)}_{(q+k)^\perp,0}\chi_{k^\perp}\mathbf{z}^{-\frac{1}{2}(k^\perp,k^\perp)}\frac{1}{\hat{k}^2}\delta_{\mu\nu} \tag{45}$$

where $k = k^{\mathrm{ph}} + k^\perp$,

$$\chi_{k^\perp} = 1 - \delta_{k^\perp,0} \qquad , \qquad \hat{k}^2 = 4\sum_\mu \sin^2(\frac{k_\mu}{2}) \tag{46}$$

and the bilinear $(\tilde{k}^\perp, k^\perp)$ is defined as

$$(\tilde{k}^\perp, k^\perp) = \tilde{n}_x n_x + \tilde{n}_y n_y + (\tilde{n}_x + \tilde{n}_y)(n_x + n_y) \quad . \tag{47}$$

The comparison of equation (43) with the ordinary Fourier and colour decomposition of the gauge field $A_\mu(r)$, shows that the colour degrees of freedom of $A_\mu(r)$ are transferred into the $k^\perp$ (unphysical momentum) degrees of freedom: their number is just $N^2 - 1$, due to the trace condition (44).

We now specialize to our particular case, SU(2), with the twist matrices $\Omega_x$, $\Omega_y$ given by equation (14).

[2]In this gauge, the ghost propagator in momentum space has the same form as the gluon one, with $\delta_{\mu\nu}$ replaced by 1.

It follows that:

$$\begin{aligned} \Gamma_k|_{(n_x^\perp=0,n_y^\perp=1)} &= -i\sigma_1 \\ \Gamma_k|_{(n_x^\perp=1,n_y^\perp=0)} &= +i\sigma_2 \\ \Gamma_k|_{(n_x^\perp=1,n_y^\perp=1)} &= -i\sigma_3. \end{aligned} \tag{48}$$

By rewriting equation (43) in the form

$$A_\mu(r) = \frac{1}{L^4}\sum_{k^{\rm ph},k^\perp} \Gamma_{k^\perp}\tilde{A}_\mu(k^{\rm ph},k^\perp)\mathrm{e}^{ikr+i\frac{k_\mu}{2}} = \frac{1}{L^4}\sum_{k^{\rm ph},c}\sigma_c\tilde{A}^c_\mu(k^{\rm ph})\mathrm{e}^{ikr+i\frac{k_\mu}{2}}, \tag{49}$$

we find:

$$\begin{aligned} \tilde{A}^1_\mu(k^{\rm ph}) &= -i\tilde{A}_\mu(k^{\rm ph},k^\perp)|_{(n_x^\perp=0,n_y^\perp=1)} \\ \tilde{A}^2_\mu(k^{\rm ph}) &= +i\tilde{A}_\mu(k^{\rm ph},k^\perp)|_{(n_x^\perp=1,n_y^\perp=0)} \\ \tilde{A}^3_\mu(k^{\rm ph}) &= -i\tilde{A}_\mu(k^{\rm ph},k^\perp)|_{(n_x^\perp=1,n_y^\perp=1)}, \end{aligned} \tag{50}$$

where a correspondence between the $k^\perp$ and the colour $c$ indices is explicitly established. The details of this correspondence depend on the choice for the $\Omega$ matrices.

The gluon propagator (45) becomes in this colour basis and for the SU(2) case :

$$\langle\tilde{A}^a_\mu(q^{\rm ph})\tilde{A}^b_\nu(k^{\rm ph})\rangle = \frac{1}{4}\delta^{(4)}_{(q+k)^{\rm ph},0}\delta^{ab}\frac{1}{\hat{k}^2}\delta_{\mu\nu} = \tilde{G}^a_{\mu\nu}(k^{\rm ph}) \tag{51}$$

An ordinary Fourier transform gives the corresponding propagator in configuration space (which is diagonal in colour indices):

$$G^c_{\mu\nu}(r) = \frac{1}{L^4}\sum_{k^{\rm ph}}\tilde{G}^c_{\mu\nu}(k^{\rm ph})\mathrm{e}^{ik^{\rm ph}r+ik^\perp(c)r} \tag{52}$$

where $k^\perp(c)$ is given by the correspondence established in equation (50). The phase factor containing $k^\perp(c)$ is responsible for the peculiar, colour dependent properties of $G^c_{\mu\nu}(r)$ under translations by $L$ in the twisted directions.

# Acknowledgements

The idea of using twisted Polyakov loops for a definition of $\alpha_s$ was suggested to us by M. Lüscher. We thank him and P. Weisz for useful discussions and for providing us with notes on Feynman rules in momentum space.

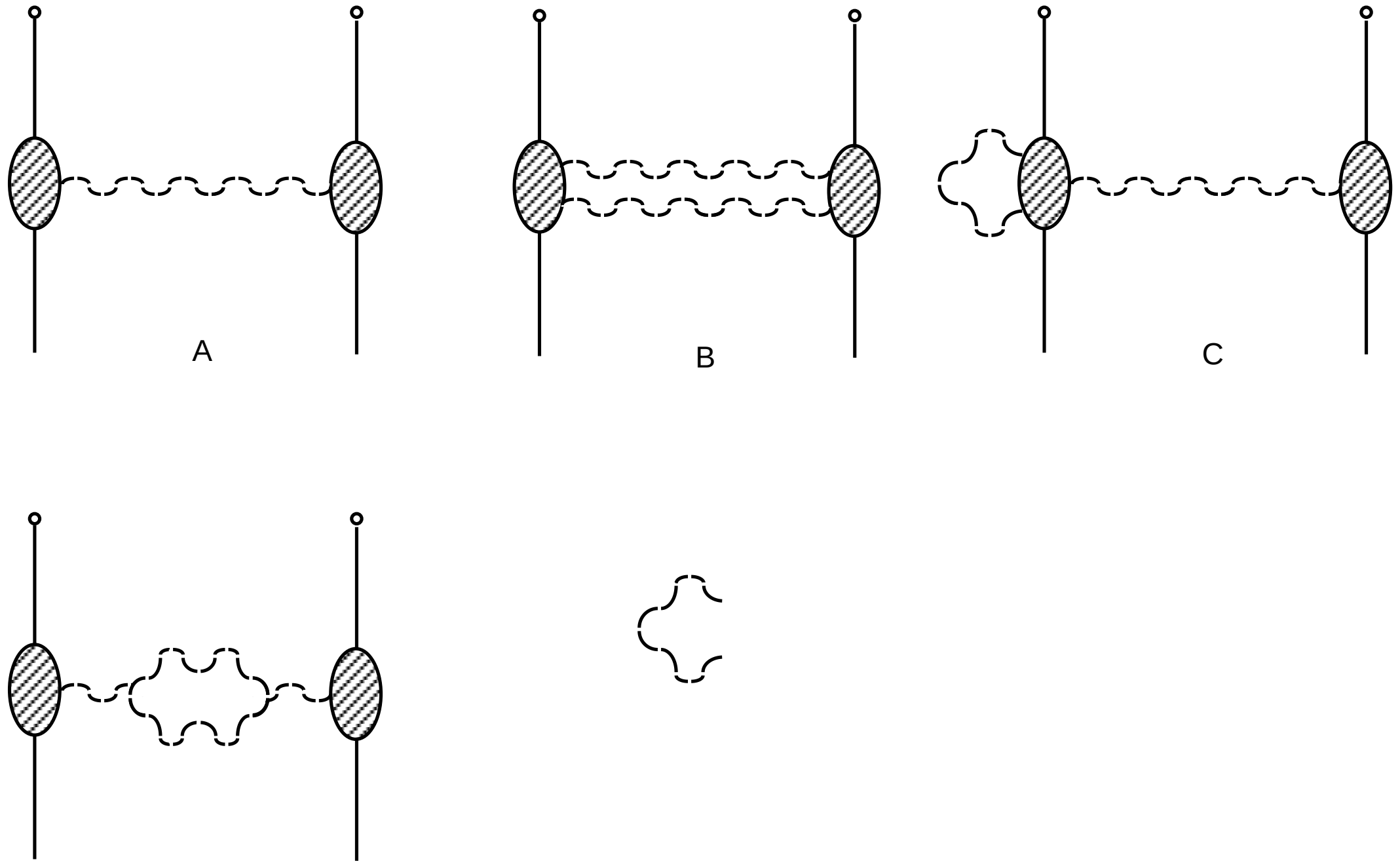
A
B
C